\documentclass
[12pt]
{article}
\begin{document}

\newcommand{\bean}{\begin{eqnarray*}}
\newcommand{\eean}{\end{eqnarray*}}
\newcommand{\ed}{\end{document}}
\newcommand{\pr}{\prime}
\newcommand{\ppr}{\prime\prime}
\newcommand{\cE}{{\tilde E}}
\newcommand{\vphi}{{\varphi}}
\newcommand{\oO}{O(k^{-1})}
\newcommand{\be}{\begin{equation}}
\newcommand{\ee}{\end{equation}}
\newcommand{\barr}{\begin{array}}
\newcommand{\earr}{\end{array}}
\newcommand{\bea}{\begin{eqnarray}}
\newcommand{\eea}{\end{eqnarray}}
\newcommand{\pa}{\partial}
\newcommand{\xx}{\hbox{}^*_*}
\newcommand{\sds}{\subset\hskip - 1em +}

\title{Realizations of observables in
Hamiltonian systems
with first class constraints.
} \author{A.V.Bratchikov \\ Kuban State
Technological University,\\ 2 Moskovskaya Street, Krasnodar, 350072,
Russia\\ E-mail:bratchikov@kubstu.ru} \date {December,\,2004} \maketitle

\begin{abstract}
In a Hamiltonian system with first class
constraints
observables
can be defined as elements of a quotient Poisson
bracket algebra. In the gauge fixing method
observables form a quotient Dirac bracket algebra.
We  show that these two algebras are isomorphic.
A new realization of the observable algebras through the original
Poisson bracket is found.Generators,
brackets and pointwise products of the  algebras under
consideration are calculated.
\end {abstract}
\bigskip


\bigskip

\newpage
{\bf1.}
In a Hamiltonian system with the first class
constraints $\varphi_j(p,q),j=1\ldots J,$
\bea
\label{p}
\left. [\varphi_i,\varphi_j]\right
|_{\varphi=0}=0,
\eea
physical functions are elements of a Poisson bracket
algebra of the first class functions
\bea \label{phy}
P =\{f(p,q)\,|\,\left. [f,{\varphi_j}]
\right|_{\varphi=0}=0
\}.
\eea
Here $p=(p_1,\ldots, p_n) ,q=(q_1,\ldots, q_n)$ are the canonical
coordinates.
Observables are elements of the algebra $P/I,$
where
\bean \label{phyi}
I =\{u(p,q)\,|\,\left. u\right|_{\varphi=0}=0\}
\eean
(see e.g. \cite{GT} and references therein).
These definitions correspond to the Dirac quantization \cite {D1}
without gauge fixing.

In the gauge fixing method \cite {D2} the gauge functions
$\chi_i(p,q), i=1\ldots J,$ are
introduced which serve as auxiliary constraints. $\chi_i$
are supposed to satisfy the conditions
\bean \left.\det ([\chi_i,\varphi_{j} ])\right|_{\varphi=\chi=0} \ne 0  \eean
and the constraints
$(\pi_\alpha )=(\varphi_{1},\ldots,\varphi_{J},\chi_1,\ldots \chi_J)$
are second class.
Then
the original Poisson bracket is replaced by the Dirac  one
\bean
\label{Dbr} [g,h]_D=[g,h]- [g,\pi_{\alpha}]c_{\alpha
\beta}[\pi_{\beta},h],\qquad c_{\alpha
\beta}[\pi_{\beta},\pi_{\gamma}]=\delta_{\alpha \gamma} .  \eean
The constraints  $(\pi_\alpha)$
are first class with respect to the Dirac bracket  and
physical functions are defined by the equations
\bean \label{hjkl}
\left.{[f,\pi_\alpha]_D}\right|_{\pi=0}=0
\eean
which are satisfied identically.
Let $A$ be the
space of all the functions on phase space
and
\bean \label{phyi}
\Phi =\{v(p,q) \,| \,\left. v \right|_{\pi=0}=0\}.
\eean
The algebra of
observables in the gauge fixing method is the Dirac bracket algebra
$A/\Phi.$
In fact,$A/\Phi$
is a
Poisson algebra, as well as $P/I.$
A connection between the classical Hamilton equations which correspond
to these two methods is described in \cite {GT}.

In the present paper we show that $P/I$ and $A/\Phi$
are isomorphic as Poisson algebras.

In a recent article \cite {Br} a family of the new algebras with respect
to the original Poisson bracket was constructed which are isomorphic
to a Dirac bracket algebra. The algebras which are isomorphic to
$A/\Phi$ give us realizations of $P/I.$

Using gauge fixing functions we find a new realization of $P/I.$
It looks like $Q/K,$ where  $Q$ and $K$ are Poisson subalgebras of $P$
and  $I$ respectively.

To describe elements of  $P/I$  and
prove the isomorphisms we find a local solution to equations
(\ref {phy}). Similar equations determine elements of $Q.$
Explicite expressions for generators enable us to
calculate brackets and pointwise products  for the observable algebras
$P/I$ and $Q/K.$

We shall assume that all the quantities vanishing on a
constraint surface are linear functions of the constraints.

\bigskip

{\bf 2.}
To find elements of $P$ explicitly
let us consider  the equations
\bea
\label {Phy}
[f,{\varphi_j}]
\in I.
\eea
with the initial condition
\bea
\label{in}
f(p,q) \in  \{f_0(p,q)\}.
 \eea
Here $\{f_0\}\in A/\Phi$ is the coset represented by
$f_0 \in A.$

Due to (\ref{in})
\bea
\label{ins}
f = f_0+ r_i\varphi_i+ s_j\chi_j
\eea
for some $r_i=r_i(p,q),s_j=s_j(p,q).$
Substituting this
into equation (\ref {Phy}),
we get
\bean
[f_0,\varphi_i]+\chi_j[s_j,\varphi_i]+s_j[\chi_j,\varphi_i] \in I
\eean
or
\bean
\psi_k+(Bs)_k \in I.
\eean
Here
\bean
\psi_k=[f_0,\varphi_i]b_{ik},\quad
 (Bs)_k=s_k+\chi_j[s_j,\varphi_i]b_{ik},\quad
[\chi_i,\varphi_j]b_{jk}=\delta_{ik}.
\eean
We assume that the operator  $B$
is locally invertible.
For $u_i \in I$ we have $(B^{-1}u)_i \in I.$ Hence
\bea
\label {phyn}
s_j =-(B^{-1}\psi)_j+s_{jk}\varphi_k
\eea
for some functions $s_{jk}(p,q).$
Expressions (\ref {ins},
\ref {phyn}) give us  a solution to equations (\ref
{Phy}) with initial condition (\ref {in}).

We have shown that
for any $f_0 \in A$ the set
$\{f_0\}\cap P$, consists of all the
expressions
\bean \label{geg3}
f = L(f_0)
+ r_{i} \varphi_{i}.
\eean
Here $r_{i}(p,q)$ are arbitrary functions and
\bean \label{geg3}
L(f_0) =f_0 -\chi_j(B^{-1}\psi)_j. \eean
From this it follows
\bea
\label{geg33}
 \{f_0\}\cap P=\{L(f_0)\}_{P/I}.
\eea
Here $\{L(f_0)\}_{P/I} \in P/I$ denotes the coset represented by
$L(f_0) \in P.$

To show that $P /I$ is isomorphic
to
$A/\Phi$
let us define the linear function
$T:
P /I\to A /\Phi$
\bea  \label {T}
T(\{f\}_{P/I})=\{f\} .
\eea
Due to (\ref {geg33}) the inverse function $T^{-1}:A /\Phi \to
P /I$ is given by \bean \label {T-1}
T^{-1}(\{f_0\})=\{L(f_0)\}_{P/I} .\eean

To show that $T$ is a homomorphism let us  compute
\bean \label
{mom2} T([\{f\}_{P/I},\{g\}_{P/I}])= T(\{[f,g]\}_{P/I}).
\eean
Due to (\ref {p}) for $f,g\in P$
\bean \label
{mom23} [f,g]- [f,g]_D  \in I.
\eean
From this and definition (\ref {T}) it follows
\bean
T(\{[f,g]\}_{P/I})=
T(\{[f,g]_D\}_{P/I})=\{[f,g]_D\}= [\{f\},\{g\}]_D=
[T(\{f\}_{P/I}),T(\{g\}_{P/I})]_D.
\eean
Hence the Dirac bracket algebra
$A/\Phi$ is isomorphic to the Poisson
bracket algebra $P /I.$

It is easy to check that $T$ is also an isomorphism with respect to the
pointwise multiplication.We get \bean
T(\{f\}_{P/I}\{g\}_{P/I})= T(\{fg\}_{P/I})=\{fg\}= \{f\}\{g\}=
T(\{f\}_{P/I})T(\{g\}_{P/I}).
\eean
Thus we have shown that
$A/\Phi$ and  $P /I$ are
isomorphic as Poisson
algebras.

\bigskip

{\bf 3.}
Let us define the
space
\bea
\label{phyii}
Q=\{F\in P\,|\,
[{\chi_j},F]\in I\}.
\eea
One can check that $Q$ is a Poisson algebra
and $K=Q\cap I$ is an ideal of $Q.$

Let $F$ be a solution to equations (\ref {phyii}) with the initial
condition
\bean
F(p,q)\in \{F_0(p,q)\}_{P /I}.
\eean
The function $F$ can be represented in the form
\bea \label {phyk} F= F_0+
\nu_{i}\varphi_{i}, \eea
for some  $\nu_{i}=\nu_{i}(p,q).$
Substituting
(\ref {phyk})
into equations
(\ref {phyii})
we get
\bean  \label{ge}
[\chi_{j},F_0]+ \nu_{i}
[{\chi_j},\varphi_{i}]\in I.
\eean
A solution to these equations is
\bean  \label{ges}
\nu_{i}=- b_{ij}[\chi_{j},F_0]
+\nu_{ij}\varphi_{j}.
\eean
Here $\nu_{ij}=\nu_{ij}(p,q)$  are arbitrary functions.

We have proven  that
for any $F_0 \in P$ the set
$\{F_0\}_{P/I}\cap Q$ consists of all the
expressions
\bean
\label{geg39}
F=R(F_0)+
\nu_{ij} \varphi_{i}\varphi_{j},
\eean
where
\bean
R(F_0)= F_0-b_{ij}[\chi_{j},F_0]\varphi_{i}
\eean
From this it follows
\bea
\label{geg37}
 \{F_0\}_{P/I}\cap Q=\{R(F_0)\}_{Q/K}.
\eea
Here
$\{R(F_0)\}_{Q/K} \in Q/K$ is the coset represented by
$R(F_0)\in Q.$

Our aim is to show that $Q /K$ is isomorphic to  $P /I.$
Let us define the linear function
$S:
Q/K \to P /I$
\bean  \label {S}
S(\{F\}_{Q /K})=\{F\}_{P/I} .
\eean
Due to equation (\ref {geg37}) the inverse function $S^{-1}:P /I \to
Q /K$ is given by \bean \label {S-1}
S^{-1}(\{F_0\}_{P/I})=\{R(F_0)\}_{Q /K} .\eean

To show that $S$ is a homomorphism we have the following
computations
\bean
S(\{[F,G]\}_{Q /K})=
\{[F,G]\}_{P/I}= [\{F\}_{P /I},\{G\}_{P /I}]=
[S(\{F\}_{Q /K}),S(\{G\}_{Q /K})].
\eean
Hence $Q /K$ and  $P /I$ are isomorphic as Poisson bracket algebras.
It is easy to check that $S$ is also a homomorphism  with respect
the pointwise multiplication
\bean
S(\{F\}_{Q /K}\{G\}_{Q /K})=
S(\{F\}_{Q /K})S(\{G\}_{Q /K}).
\eean
This tells us that $Q /K$ and  $P /I$ are isomorphic as Poisson
algebras.

Brackets and pointwise products for observables can be
calculated as follows.
One can check that for $f=L(f_0)\in P$ and $g=L(g_0)\in P$ the
functions $[f,g]$ and $fg$ satisfy equations (\ref {Phy}) with the
initial conditions $[f,g]\in \{[f_0,g_0]_D\}$ and $fg \in  \{f_0g_0\}$
respectively. Due to
(\ref{geg3})
\bean \label {cr5} [f,g]=L([f_0,g_0]_D)+\tilde
u,\quad fg=L(f_0g_0)+\tilde w,
\eean
where $\tilde u,\tilde w \in I.$
From this it follows
\bean \label {cr57}
[\{f\}_{P/I},\{g\}_{P/I}]=\{L([f_0,g_0]_D)\}_{P/I},
\eean
\bean
\{f\}_{P/I}\{g\}_{P/I}=\{L(f_0g_0)\}_{P/I}.  \eean

Consider the algebra $Q/K.$
For $F=R(F_0)\in Q$ and $G=R(G_0)\in Q$ the
functions $[F,G]$ and $FG$ satisfy equations (\ref {phyii}) with the
initial conditions $[F,G]\in \{[F_0,G_0]\}_{P/I}$ and $FG \in
\{F_0G_0\}_{P/I}$ respectively.  According to (\ref {geg37}) \bean
\label {cr58} [F,G]=R([F_0,G_0])+\tilde U,\quad FG=R(F_0G_0)+\tilde W,
\eean
where $\tilde U,\tilde W\in K.$
Therefore, we have
\bean \label {cr57}
[\{F\}_{Q/K},\{G\}_{Q/K}]=\{R([F_0,G_0])\}_{Q/K},
\eean
\bean
\{F\}_{Q/K}\{G\}_{Q/K}=\{R(F_0G_0)\}_{Q/K}.  \eean

\bigskip

{\bf Acknowledgements}

The research was suppoted in part by RFBR grant 03-02-96521.

\end{document}